\documentclass[conference]{IEEEtran}
\usepackage{amsmath, amssymb, amsfonts}
\usepackage{subcaption}

 \usepackage{algorithmic}
\usepackage{graphicx}
\usepackage{multirow}
\usepackage{xcolor}
\usepackage{array}

\def\BibTeX{{\rm B\kern-.05em{\sc i\kern-.025em b}\kern-.08em
    T\kern-.1667em\lower.7ex\hbox{E}\kern-.125emX}}
\begin{document}

\title{Propeller Modulation Equalization via Reference Tones}


\author{\IEEEauthorblockN{Mostafa Ibrahim}
\IEEEauthorblockA{\textit{Department of Engineering Technology} \\
\textit{Texas A\&M University}\\
Texas, USA \\
mostafa.ibrahim@tamu.edu}
\and
\IEEEauthorblockN{Sabit Ekin}
\IEEEauthorblockA{\textit{Departments of Engineering Technology, } \\
\textit{and Electrical \& Computer Engineering}\\
\textit{ Texas A\&M University}\\
Texas, USA \\
sabitekin@tamu.edu}
}

\maketitle 
\begin{abstract}
Propeller modulation, also known as micro-Doppler modulation, presents a significant challenge in radio frequency (RF) inspection operations conducted via drones. This paper investigates the equalization of propeller modulation effects on RF signals, specifically targeting applications in navigation aids such as Instrument Landing Systems (ILS). By employing a continuous reference tone, the propeller-induced Doppler spread can be effectively captured and equalized, improving signal integrity and accuracy. Simulation results demonstrate that the proposed equalization method significantly reduces DDM deviation caused by propeller modulation, even under various propeller speeds. The findings suggest that incorporating such equalization techniques can enhance the reliability and efficiency of drone-based RF inspections.
\end{abstract}


\IEEEpeerreviewmaketitle

\section{Introduction}
This paper focuses on equalizing propeller modulation in radio frequency (RF) inspection operations conducted via drones. The future of RF inspections for Federal Aviation Administration (FAA) services, such as Instrument Landing System (ILS) and VHF Omnidirectional Range (VOR) or radar services, is projected to rely on unmanned aerial vehicles (UAVs), which utilize propellers for hovering. These drones can be used for navigation aids (NAVAIDS) inspection, as reported by the International Civil Aviation Organization (ICAO), and several companies already offer this service. Additionally, drones are proposed for radar calibration \cite{boyer2017uav,yin2019uav} and antenna pattern measurements \cite{salazar2023accuracy,garcia2017antenna,segales2023far,culotta2021uncertainty,fernandez2018use,salari2021unmanned}.

Propeller modulation, also known as micro-Doppler propeller modulation, is a channel disruptor caused either by cutting the direct signal path or altering the reflective environment around the receiver attached to the drone. This effect has been characterized in the literature through simulations, measurements, and mathematical models in several studies \cite{marak2017electromagnetic,faul2021impact}. Propeller modulation has also been utilized for drone recognition and classification \cite{mingjiu2022rotor,chen2019signal,ezuma2022comparative,maasdorp2015fm,pan2013modulation,liu2017micro,yong2016research}, and specifically for drones and propeller parameter extraction \cite{kang20206}. However, these studies primarily focus on radar theory and do not address the impairment in communication or inspection systems.

The impairments caused by propeller modulation were reported in the Scanning Beam Microwave Landing System \cite{pope1974propeller}. The Pilot’s Handbook of Aeronautical Knowledge \cite{federal2009pilot} reports that specific propeller RPM settings or helicopter rotor speeds can affect the VOR course deviation indicator. Additionally, certain engine RPM settings in propeller-powered planes can impact the auto-pilot system that relies on ILS measurements \cite{duncanaviation2019propeller}. This aligns with studies on ILS inspection that identify propeller modulation as a measurement noise \cite{horapong2017design,ibrahim2022unmanned,8448050,bredemeyer2018employing}. In ILS, for instance, the propeller causes interference between the center tone and the 90Hz and 150Hz tones.

This paper investigates the possibility of equalizing the propeller modulation effect for communication purposes and radio frequency inspections. We argue that equalization is more critical for the latter, as modern communications use wide bandwidth bursts, and subcarrier spacing can be adjusted to accommodate Doppler spreads. However, NAVAIDS and radar signals, such as ILS, VOR, and frequency modulated continuous wave (FMCW) chirps, are continuous and require equalization to obtain accurate inspection data. Without equalization, averaging would be necessary to manage the time variation of the channel, resulting in longer inspection times.

The novelty of this work lies in proposing a continuous reference tone for equalizing the propeller modulation effect on RF signals, specifically targeting applications in NAVAIDS inspections. By employing a continuous reference signal that matches the characteristics of the inspected signal, we aim to capture and equalize the Doppler spread, thereby improving signal integrity and accuracy.

The rest of this paper is organized as follows: Section II describes the system model and the representation of propeller modulation. Section III details the equalization formulation using the proposed continuous reference tone. Section IV presents the simulation results and discussion. Finally, Section V concludes the paper and suggests directions for future research.

\section{System Model}
\begin{figure}[h]
    \centering
    \includegraphics[width=.46\textwidth]{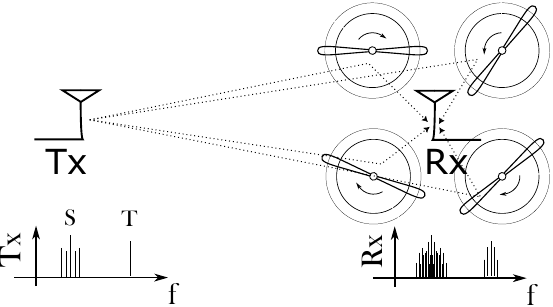}
    \caption{System Model representation.}
    \label{fig:sysmod}
\end{figure}
\subsection{Propeller Modulation Representation}

Propeller modulation has been characterized and modeled in several studies. It is physically resulting from a chopping effect that modulates the received signal with a cyclic process with an interval related to a full revolution of the propeller blades. The modulating signal is shaped in relationship to the propeller shapes e.g. their lengths and number of blades.  We generalize this effect as the cyclic function $g(\omega t-\phi)$, where $\omega$ is related to the speed of the propeller, and $\phi$ is its phase. The process $g$ can resemble a square wave, a clipped cosine, or any cyclic shape. A transmitted signal $S(t)$ is received at the inspection drone as:
\begin{equation}
    R(t) = \sum_{p=1}^N a_p~S(t-\tau_p)~\sum_{i=1}^M a_i~\exp(j\omega_i t),
\end{equation}
where $N$ is the number of propellers, and each one has a different speed and phase, and $a_p$ and $a_i$ are modulation coefficients. $M$ is the number of discrete Fourier transformed tones representing the micro-Doppler modulation. The values $a_p$ are related to how the propellers are positioned in the propagation path. The delay value $\tau_p$ corresponds to the different path lengths of the signal in relation to each blade. However, considering the small distance between the different propellers, the delay will be in sub-nano seconds, and that the inspected signal is of narrow bandwidth (in the range of kiloHertz), the factor $\tau_p$ can be dropped. 

Regardless of the values of $a_p, a_i,~\omega_p,~N,$ and $M$, the total propeller modulation effect can be captured and equalized using a proper reference signal/tone. In the next subsection, we propose the continuous tone as a reference signal and compare it conceptually with the reference signals used in communication systems
\begin{figure}[h]
    \centering
    \includegraphics[width=.48\textwidth]{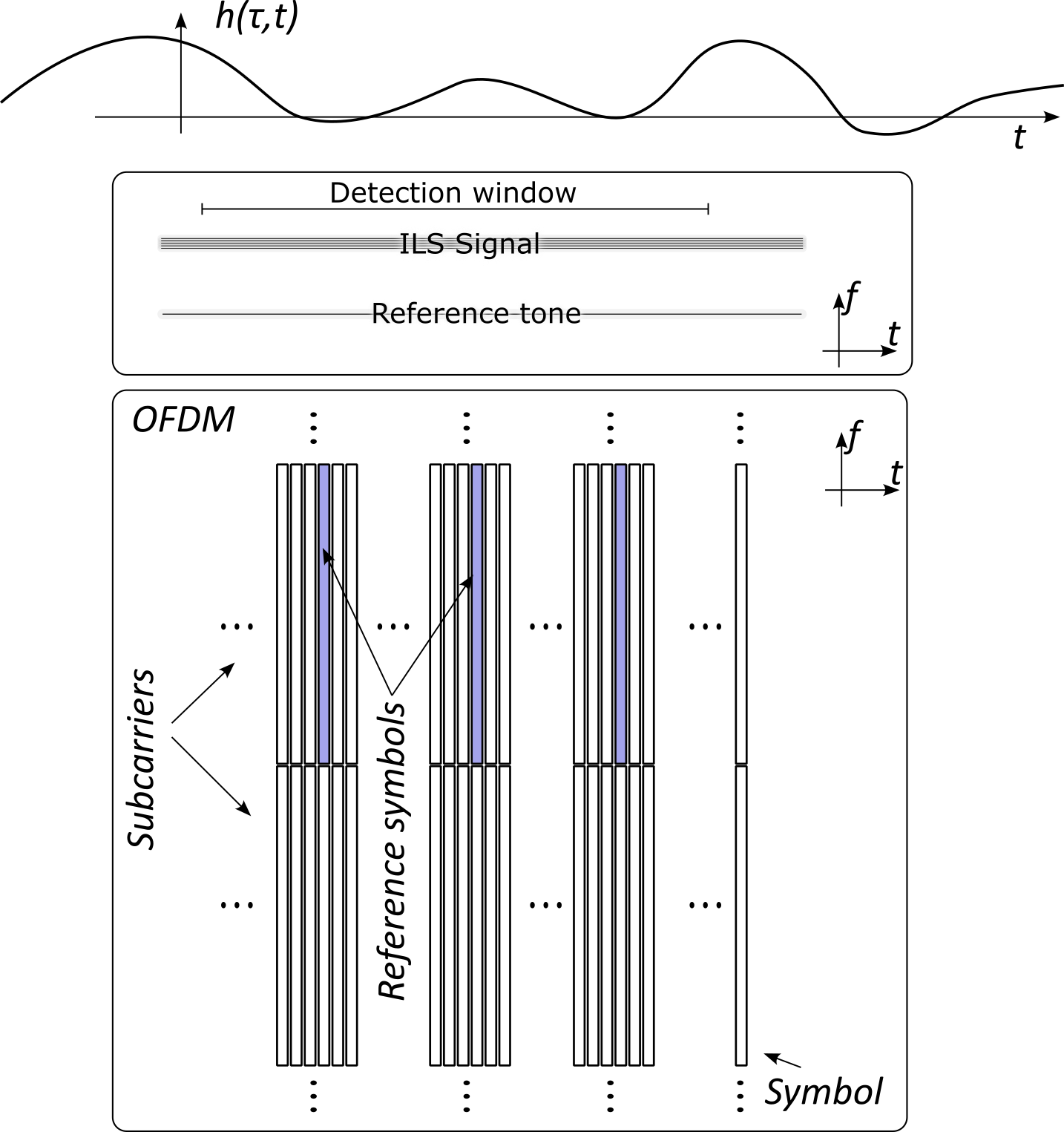}
    \caption{Reference signals comparison between continuous signal inspection and modern communication systems.}
    \label{fig:tonevsptrs}
\end{figure}
\subsection{Reference Tone}
Doppler spread in modern communication systems, such as 4G and 5G, can be mitigated by adjusting subcarrier spacing and symbol duration relative to the channel's coherence time. Subcarrier spacing is typically chosen to be wider than the Doppler frequency spread, and symbol durations are kept shorter than the coherence time of the channel. This approach ensures that the channel variations within one symbol period are minimal.

Communication systems utilize reference signals, or pilots, to estimate and track temporal variations of the channel. In Orthogonal Frequency Division Multiplexing (OFDM)-based systems like 4G and 5G, these pilots are embedded in the resource grid. The short duration of OFDM symbols, compared to the channel coherence time in urban environments, allows pilots to be sparse in time, often repeating every few data resource blocks.

In contrast, navigation aid signals, such as those used in Instrument Landing Systems (ILS) and VHF Omnidirectional Range (VOR), have much lower bandwidths, typically ranging from a few Hz to a few kHz. Consequently, the receiver's capturing window for these signals is significantly longer than that for bursty communication systems. Moreover, navigation aid signals are continuous, unlike the burst transmissions in modern communications, which means that synchronization of the reception window is not required. This simplicity allows the receiver to operate without the need for complex synchronization mechanisms.

To effectively deal with the Doppler spread in these continuous, low-bandwidth signals, a continuous reference signal is necessary. This continuous reference signal must match the characteristics of the inspected signal to accurately capture temporal variations within the receiving window. The presence of a continuous reference signal ensures that the channel variations can be tracked consistently over the entire duration of the signal capture, providing a stable basis for equalization.

In Fig. \ref{fig:tonevsptrs}, we illustrate a comparison between the resource grid of a communication system and that of a navigation system, highlighting their interaction with the temporal channel variation of a channel tap at delay $\tau$. This figure demonstrates how the continuous reference signals in navigation systems are essential for maintaining the accuracy of signal measurements and ensuring reliable performance in the presence of Doppler effects.

\section{Equalization Formulation}
The proposed continuous tone is used to determine the Doppler spread. The equalization process determines the modulation signal and divides it into the time domain. The equalization is calculated over a captured duration in time, over which Fourier and inverse Fourier transforms are used to transform the signal back and forth between time and frequency domains.  

A signal $S(t)$ is accompanied by an orthogonal tone $T(t)$ with a frequency separation wider than twice the expected Doppler spread. The received signal is:
\begin{equation}
    R(t)= \sum_{p=1}^N ~ a_p~[S(t)+ T(t) ]~g(\omega_p t-\phi_p).
\end{equation}
Due to the separation between the reference tone and the signal, we can write the frequency domain as:
\begin{equation}
    \mathbb{R}(f)= \mathbb{S}(f) \otimes\mathbb{G}(f) + \mathbb{T}(f) \otimes\mathbb{G}(f) + N(f),
\end{equation}
where the transformed $\mathbb{G}(f)= \int_{0}^{T_c} \sum_{p=1}^N a_p g(\omega_p t-\phi_p) e^{j 2\pi f t} dt.$ $T_c$ is the captured window duration. Fast Fourier Transform (FFT) is to be used to realize $\mathbb{G}(f)$ 
\subsection*{Extracting Doppler modulation process:}
The two received signal components can be separated via windowing processes. The Doppler modulation process can be found from:
\begin{equation}
    \mathbb{D}(f)=\mathbb{R}(f). \mathbb{W}^{(T)}(f),
    \label{eq:eq4}
\end{equation} 
where $\mathbb{W}^{(T)}$ is a band pass window passing only the tone band. Hence
\begin{equation}
    \mathbb{D}(f)=\mathbb{T}(f) \otimes\mathbb{G}(f) + \mathbb{N}^{(T)}(f). 
\end{equation} 
The doppler modulation process in time domain is realized from:
\begin{equation}
    D(t)= IFFT \left( \mathbb{D}(f)\right)
\end{equation}
The equalized signal is 
\begin{equation}
    S(t)=\dfrac{IFFT\left(\mathbb{R}(f). \mathbb{W}^{(S)}(f)  \right)}{IFFT \left( \mathbb{D}(f)\right)},
    \label{eq:equalization}
\end{equation}
where $\mathbb{W}^{(S)}$ is a band pass window passing only the signal band and rejecting the reference tone.


    \label{fig:denoise}

\section{Simulation and Discussion}

In this section, we evaluate the equalization process assuming the signal under propeller Doppler spread is an Instrument Landing System (ILS) signal. An ILS signal is a navigational aid used in aviation that provides guidance for the final approach to a runway. It consists of a carrier and modulations at $\pm 90~Hz$ and $\pm 150~Hz$. These side carriers are transmitted over offset beams to provide position information relative to the carrier amplitudes. The position is calculated from the difference in depth of modulation (DDM) as:
\begin{equation}
    DDM = \dfrac{A_{90} - A_c}{A_c} - \dfrac{A_{150} - A_c}{A_c},
    \label{eq:DDM}
\end{equation}
where $A_c$, $A_{90}$, and $A_{150}$ are the amplitudes of the carrier, 90 Hz, and 150 Hz modulations, respectively.

\begin{figure}[t]
    \centering
    \begin{subfigure}[b]{0.48\textwidth}
        \centering
        \includegraphics[width=\textwidth]{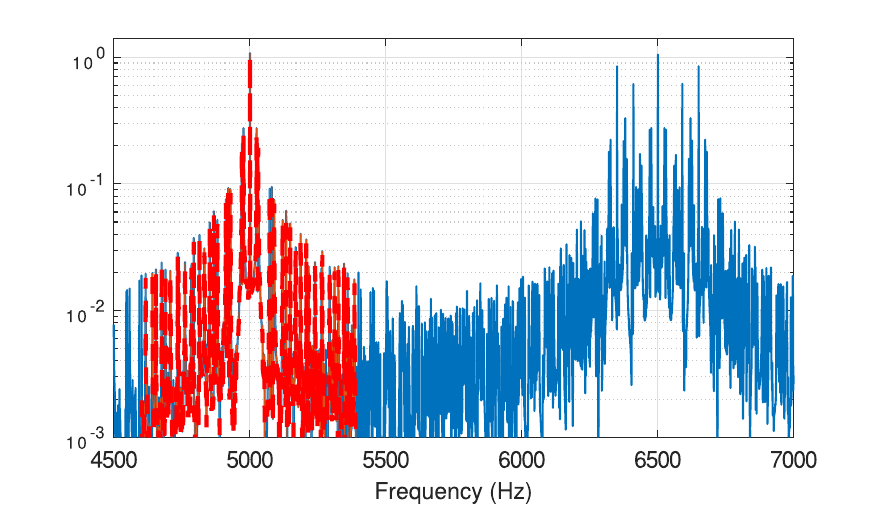}
        \caption{Propeller modulated $\left( \mathbb{S}(f) + \mathbb{T}(f) \right)$.}
        \label{fig:simmodulated}
    \end{subfigure}
    \begin{subfigure}[b]{0.48\textwidth}
        \centering
        \includegraphics[width=\textwidth]{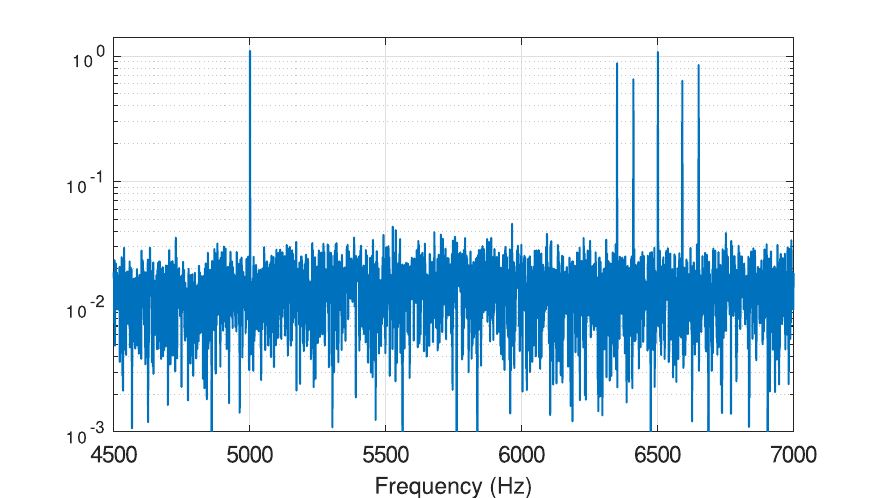}
        \caption{Equalized $\left( \mathbb{S}(f) + \mathbb{T}(f) \right)$.}
        \label{fig:simequalized}
    \end{subfigure}
    \caption{Comparison of Propeller Modulated and Equalized $\left( \mathbb{S}(f) + \mathbb{T}(f) \right)$.}
    \label{fig:comparison}
\end{figure}

In the presence of propeller modulation, there is interference between the components of the ILS signal. To resolve this, we transmit the pilot tone 1.5 KHz away from the ILS signal, as shown in Fig. \ref{fig:simmodulated}. The figure shows the modulation effect in the frequency domain on the combined signal $\mathbb{S}(f) + \mathbb{T}(f)$.

The propeller modulation used in this simulation is formulated as
\begin{equation}
    R(t) =  S(t) \cdot g(2 \pi f_p t),
\end{equation}
where $g$ is a square wave, which gives a harsh modulation effect due to its sinc-shaped frequency domain representation. We capture $\pm 300~Hz$ of the modulated tone and use it to equalize the signal using Eq. (\ref{eq:equalization}). The equalization results are shown in Fig. \ref{fig:simequalized}. We notice that the equalization results in a pure reference tone and ILS signal but creates a noise floor despite no noise being added during the modulation step. However, even with this noise floor, the signal performance is improved due to the recovered signal integrity.

The parameters of the simulation are: sampling rate = 32 KHz, capture duration = 1 second, FFT and IFFT sizes = 32000, providing a one-to-one relationship between frequency bins and frequency in Hz. The channel is modeled as additive white Gaussian noise with one tap that is Doppler modulated with $g(t)$.

\begin{figure*}[t]
    \centering
    \includegraphics[width=\textwidth]{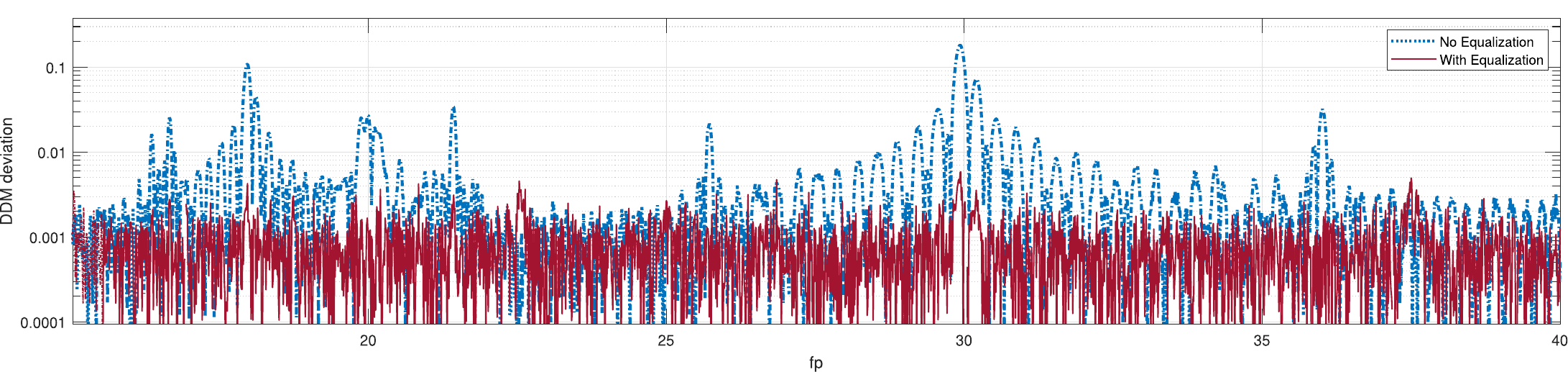}
    \caption{Comparison between the equalized and non-equalized DDM deviation at different $fp$ values, representing a sweep on propeller speeds. The simulated SNR at the receiver is 20 dB. }
    \label{fig:speed}
\end{figure*}

We evaluate the signal integrity by calculating the DDM error before and after equalization. The transmitted tone amplitudes are $A_c = 1$ V, $A_{90} = 0.6$ V, and $A_{150} = 0.8$ V, assuming the amplitude unit is in volts. The DDM value is -0.2 and is unitless. For a realistic scenario, we assume an SNR of 20 dB. The received DDM drift is calculated for different $f_p$ values, emulating different propeller speeds, and is shown in Fig. \ref{fig:speed}. The range of $f_p$ is [15, 40] Hz, passing through points where interference between the ILS tones is highest, observed at $f_p = 17, 20, 26, 30,$ and $36$ Hz. Note that  this effect is specific to the assumed modulating signal $g(t)$, which is chosen for demonstration purposes. The modulation signal will vary from drone to drone and with different propeller movements and speeds.

\begin{figure}[h]
    \centering
    \begin{subfigure}{0.48\textwidth}
        \centering
        \includegraphics[width=\textwidth]{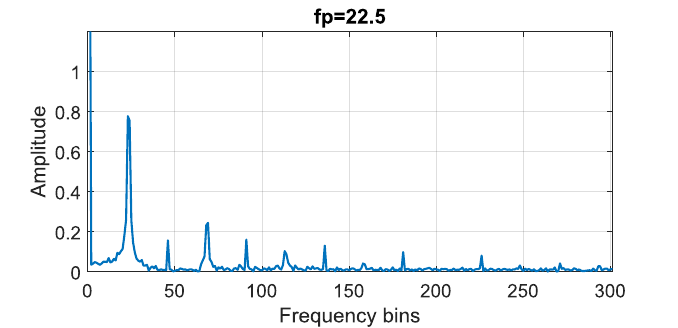}
        \caption{}
        \label{fig:blindspots1}
    \end{subfigure}
    \begin{subfigure}{0.48\textwidth}
        \centering
        \includegraphics[width=\textwidth]{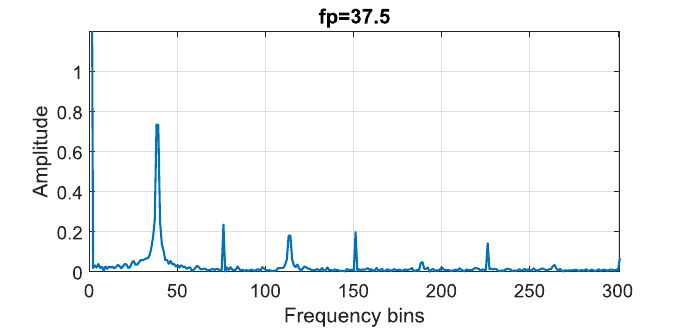}
        \caption{}
        \label{fig:blindspots2}
    \end{subfigure}
    \caption{Equalization components of $ \mathbb{G}(f) $ at two different $f_p$ values, which causes inter-carrier interference in the ILS signal after equalization.}
    \label{fig:blindspots}
\end{figure}

We observe that the equalized signal's DDM deviation is significantly lower than that of the non-equalized signal, especially at problematic propeller speeds, as the tones' interference is resolved. However, there are instances where the equalized signal shows higher deviation than the non-equalized signal, such as at $f_p = 22.5$, and 37.5 Hz, as shown in Fig. \ref{fig:speed}. This occurs because the inversion of the modulation at this $f_p$ can cause interference between the signal components. This is shown in Fig. \ref{fig:blindspots} where we plot the frequency domain representation of the equalization signal $ \mathbb{G}(f) $. In Fig. \ref{fig:blindspots1}, the equalization component at 90 Hz causes interference between the 90 Hz component of the signal and its carrier. We can observe the same effect for the 150 Hz equalization component in Fig. \ref{fig:blindspots2}. 

This effect acts as a blind spot for the equalization process at the corresponding speed values. The values presented above are specific to the simulated ILS signal, and it will occur at different speeds for different types of signals. Therefore, system designers must be aware that there will be spots in the equalization process where the non-equalized signal may be better for evaluation. However, on a broader scale of $f_p$, these points are infrequent.

\section{Conclusion}
This paper presents a novel approach to mitigating the effects of propeller modulation on RF signals using a continuous reference tone. Through simulations, we demonstrated that our equalization method significantly reduces the DDM deviation in ILS signals affected by propeller modulation.

The proposed method effectively captures and equalizes the Doppler spread caused by propeller modulation, restoring signal integrity and reducing interference. The simulation results show that equalized signals exhibit lower DDM deviation compared to non-equalized signals, particularly at problematic propeller speeds.

However, there are instances where equalization may increase signal deviation due to specific propeller speeds, but these are infrequent. Overall, the equalization process offers substantial improvements in signal accuracy for drone-based RF inspections.

Future work could explore adaptive equalization techniques tailored to varying propeller speeds and configurations, as well as real-world testing to validate the simulation results. Integrating such equalization methods holds promise for enhancing the performance and reliability of UAV-based RF inspection systems in aviation and other applications.



\bibliography{Bibliography}
\bibliographystyle{IEEEtran}

\end{document}